\documentclass[aps,prc,showpacs]{revtex4}
\usepackage{graphicx}
\usepackage{dcolumn}
\usepackage{bm}
\usepackage{longtable}
\usepackage{hyperref}
\usepackage{amsmath}
\begin{document}
\title{Subthreshold pion production within a transport description of central Au+Au collisions}
\author{Jun Hong$^{a}$ and P. Danielewicz$^{a}$} 
\affiliation{$^a$National Superconducting Cyclotron Laboratory, Michigan State
University, East Lansing, Michigan 48824, USA}
\begin{abstract}
Mapping out the equation of state (EOS) of nuclear matter is a long standing problem in nuclear physics. Recent emphasis is onto the density dependence of the symmetry energy, with experiments needing dedicated symmetry-energy observables. Towards the latter goal, we employ pBUU transport model \cite{PD-4} to simulate pion production in heavy ion collision (HIC). We find that the net pion yield tests the momentum dependence of nuclear mean field (MF). In exploring the sensitivity of pion observables to the symmetry energy at higher than normal densities, we find that our calculations of pion ratios contradict, at some level, predictions from both IBUU \cite{IBUU} and ImIQMD \cite{ImIQMD} models. We propose to employ the pion ratio in the high-energy tail of spectra in future experiments, to distinguish between different variants of high-density symmetry energy.
\end{abstract}
\maketitle
\section{Introduction}
Nuclear equation of state (EOS) relates different thermodynamical properties of nuclear matter, such as the energy or pressure of a nuclear system with density and temperature. The EOS relations are relevant for many physical processes, excitation of giant collective resonances, the dynamics in heavy ion collision (HIC) \cite{Betty-iso}, the formation of neutron stars and for their properties \cite{steiner}, etc. In the latter of the situations, a wide range of density and temperature is achieved in the course of system evolution, providing study grounds to understand the EOS. 

Nuclear matter itself stands for an infinite uniform nucleon system at some fixed ratio of neutron to proton number, with Coulomb interactions switched off. EOS for symmetric nuclear matter (SNM) has been by now significantly constrained. The zero-temperature energy minimizes at -16MeV per nucleon, at normal density of $\rho_0=0.16fm^{-3}$. The nuclear incompressibility K, which is scaled curvature of energy at normal density, has been determined to be 240MeV$\pm$20MeV, by studying excitations to the Giant Monopole Resonance (GMR) \cite{Garg-K}. However, EOS for asymmetric nuclear matter still has large uncertainties tied to the symmetry-energy term. 

Energy per nucleon in an asymmetric nuclear matter can be expanded in powers of asymmetry $\alpha$ of the system:
\begin{equation}
 \frac{E}{A}(\rho, \alpha)=\frac{E}{A}(\rho,0)+S(\rho)(\alpha^2+O(\alpha^4)),
 \end{equation}
where
\begin{equation}
\alpha=\frac{\rho_n-\rho_p}{\rho_n+\rho_p}.
\end{equation}
Here, the coefficient $S(\rho)$ is the symmetry energy. The density dependence of symmetry energy at $\rho<\rho_0$ has been constrained to some degree through various experimental measurements: of isospin diffusion, Pygmy dipole resonances, giant dipole resonances, etc. \cite{SymBetty-1,sym-Klimkiewicz,sym-Trippa}.  For $\rho>\rho_0$, on the other hand, our knowledge about the density dependence remains poor \cite{SymBrown,SymBaoan}. E.g. while some theoretical models predict that the symmetry energy keeps increasing with increasing density, some models predict the opposite. Symmetry energy impacts the composition of neutron stars and neutrino processes which rapidly cool neutron stars \cite{sym-Kubis}, neutron skin of heavy nuclei \cite{Nskin-1,Nskin-2}, etc. It is important to find a sensitive observable for experiments to constrain the behavior of symmetry energy at supranormal densities. Pions produced in HIC generally originate from higher than normal density regions, so pions might serve as a good probe of high density behavior of symmetry energy.  
 
In this paper, we mainly rely on pion observables to constrain the EOS. In section 2, we reexamine and optimize the momentum dependent nuclear mean field (MF) to describe the net pion multiplicities in Au+Au collisions at various beam energies, comparing results of our calculations to measurements of the FOPI Collaboration \cite{FOPI-1}. With the newly optimized MF, we examine predictions for the baryonic elliptic flow, encounter some difficulties in data comparisons, and suggest a possible resolution. In section 3, we examine the predictions for charged pion ratios arrived at with different variants of symmetry energy, compare our results with other theoretical predictions, and we seek observables to distinguish between different density-dependent symmetry energies, thus to provide guidance to central collision experiments.

\section{Momentum-dependent Nuclear Mean Field}

\subsection{Pion Multiplicity from pBUU}
The theoretical model used here is Boltzmann-Uehling-Uhlenbeck (BUU) transport model developed by P. Danielewicz {\it et al.} (often called pBUU) and originally formulated in \cite{PD-2}. Within the model, Boltzmann equations for the phase space distributions of different particles $f_X(\vec{p},\vec{r},t)$ are solved to describe the dynamics of nuclear collisions. 
\begin{equation}
\frac{\partial f_X}{\partial t}+\frac{\partial \epsilon_X}{\partial \vec{p}}\frac{\partial f_X}{\partial \vec{r}}-\frac{\partial \epsilon_X}{\partial \vec{r}}\frac{\partial f_X}{\partial \vec{p}}=\mathcal{K}^{<}_{X}(1\mp f_X)-\mathcal{K}^{>}_X f_X
\end{equation}
The index X above is for different species of particles, and $\epsilon_X$ is the single particle energy. In the energy range of interest here, the species accounted for are nucleons, pions, $\Delta$, N* resonances, and light (A$\leq$3) clusters. The r.h.s. factors, $\mathcal{K}^{<}$ and $\mathcal{K}^{>}$, are the feeding and removal rates, respectively, for specific momentum states. 

The single-particle energy is related to the net energy $E$ of the spin-symmetric system with:
\begin{equation}
\epsilon_X(\vec{p},\vec{r},t)=\frac{(2\pi)^3}{g_X} \frac{\delta E}{\delta f_X(\vec{p},\vec{r},t)},
\end{equation}
where $g_X$ is the spin degeneracy. In pBUU, the net energy consists of four components: a volume, surface, isospin-dependent component and a Coulomb contribution: 
\begin{equation}
E= \int e\,d\vec{r}+E_s+E_T+E_{coul}.
\end{equation}
See Ref. \cite{PD-4} for more details regarding the energy functional used in the model. The set of equations (3) is solved through a Monte-Carlo procedure \cite{PD-2}, with $f_x$ represented in terms of test particles. Unless explicitly stated, details of the calculations in this paper are such as in \cite{PD-4}.

pBUU has been successful in describing various experimental data \cite{PD-1,PD-2}. However the model has not been tested against measurements of pion multiplicity at incident energies near NN pion production threshold (e.g. 400A MeV). Fig. 1 shows net pion multiplicity obtained when using the momentum-independent and momentum-dependent MF in pBUU \cite{PD-4}, adjusted previously to different nuclear characteristics and data. Specifically Fig. 1(a) shows negative pion multiplicity and Fig. 1(b) shows positive pion multiplicity. The data represented in the figure are from the FOPI measurements of Au+Au central collisions (impact parameter $b \lesssim 2fm$) at 400A MeV, 800A MeV and 1.5A GeV \cite{FOPI-1}. As can be seen, pBUU with momentum-independent MF overestimates, by a factor of two, the measured multiplicities at all energies. With momentum-dependent MF, the calculations are consistent with data at the two higher energies, but at 400 MeV, the predicted yields are only about a half of those measured. The results of the calculations suggest that some weakening of the momentum dependence is required in order to arrive at an agreement between the pBUU results and FOPI data at the lowest of the beam energies. Other than momentum dependence, we explored potential impact of in-medium changes in the $\pi$ and $\Delta$ production rates \cite{PD-2,PD-4} consistent with detailed balance, but we found the impact of such changes within plausible range to be negligible on the final yields. 

\begin{figure}
\centering
\includegraphics[scale=1.5]{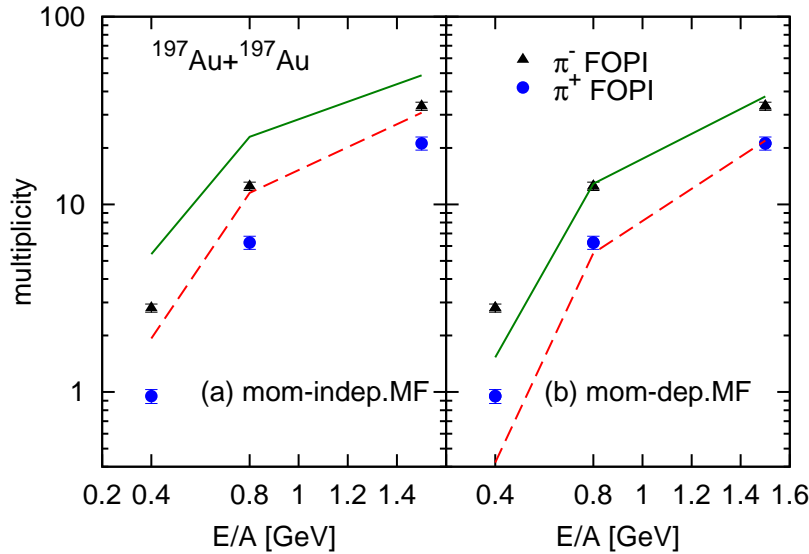}
\caption{(Color online) Pion multiplicity in central Au+Au collisions. Symbols represent data of the FOPI Collaboration \cite{FOPI-1}. The lines represent pBUU calculations when following either the momentum-independent MF (solid) or the past flow-optimized momentum-dependent MF (dashed).}
\end{figure}

The momentum dependence of MF has been implemented in pBUU through the parameterization of local particle velocity, in the following form:
\begin{equation}
v^*_X(p,\xi)=\frac{p}{\sqrt{p^2+m_X^2/(1+C\frac{m_N}{m_X}\frac{A_X \xi}{(1+\lambda p^2/m_X^2)})^2}}
\end{equation}
Here, $\xi=\frac{\rho}{\rho_0}$ and $A_X$ and $m_X$ are, respectively, the mass number and mass of species X.
The energy density of the bulk part of net energy of system is obtained from integration of the velocity over momentum, i.e. 
\begin{equation}
e=\sum\limits_X g_X \int \frac{d\vec{p}}{(2\pi)^3} \ f_X(\vec{p}) \ \left(m_X+\int_0^p  dp' \ v_X^*(p',\rho) \right)+\int_0^{\rho}  d\rho' \ U(\rho')
\end{equation}
In the equations, $U(\rho)$ represents part of the MF that only depends on density, with three parameters $a$, $b$, and $\nu$:
\begin{equation}
U(\xi)=\frac{-a\ \xi+b\ \xi^\nu}{1+(\xi/2.5)^{\nu-1}}
\end{equation}
In (6), it is the denominator in the parameterization of velocity that gives rise to the momentum-dependence of MF. In pBUU, the bulk of MF, following from (7), only acts on baryons. The momentum dependence of the MF impacts the dynamics of nucleons, thus indirectly affects pion production and pion spectra. Previously the parameterization of momentum dependent MF was adjusted using elliptic flow \cite{PD-4}. In the context of this paper, that previous parameterization will be referred to as $v_2$-optimized MF. 

We explored different possibilities for the momentum dependence of the MF by modifying the underlying parameterization for the local particle velocity. At first, we tested different density-dependencies of momentum-dependencies for the MF, by replacing the factor in (6) linear in $\xi$ by different functions of $\xi$ that reduced to 1 at $\xi \equiv \frac{\rho}{\rho_0}=1.$ However, we found the sensitivity of pion yields to that replacement to be too meager to eliminate the discrepancy between the measured and calculated pion yields. On the other hand, we found that a mere adjustment of the parameter values in (6) could reduce substantially the discrepancy between the calculated and measured net pion yields, without overly compromising the description of measured baryonic flow by the model. More discussion of that issue will come later in the paper.

In what follows, we refer to the momentum-dependent MF with the new parameters as $N_{\pi}$-adjusted MF. Parameter values for the $N_{\pi}$-adjusted and previous $v_2$-optimized MF, are listed in Table 1. From those, C and $\lambda$ dictate the momentum dependence. The net pion yields for the $N_{\pi}$-adjusted MF are displayed, together with the data, in Fig. 2.

\begin{table}[ht]
\centering
\begin{tabular}{| l | p {4cm} | p {4cm}|}
\hline
   & previous parameterization ($v_2$-optimized MF)& new parameterization ($N_{\pi}$-adjusted MF) \\
  \hline
   C &    0.643 &    0.300 \\
   $\lambda [1/c^2]$ &    0.948 &    0.400 \\
    a [MeV] &    203.92 &   173.71 \\
    b [MeV] &   65.18&    68.23 \\
   $\nu$ &    1.4838 &    1.6541 \\    
   K [MeV] &    210 &    230 \\
   m*/m &    0.7 &    0.75\\
   \hline
   \end{tabular}
   \label{table:param}
\caption{Parameters used in the previous and new momentum dependent MFs. In either case, the parameters were adjusted to yield sensible nuclear incompressibility K and nucleon effective mass m*.}
\end{table}

\begin{figure}
\centering
\includegraphics[scale=2.0]{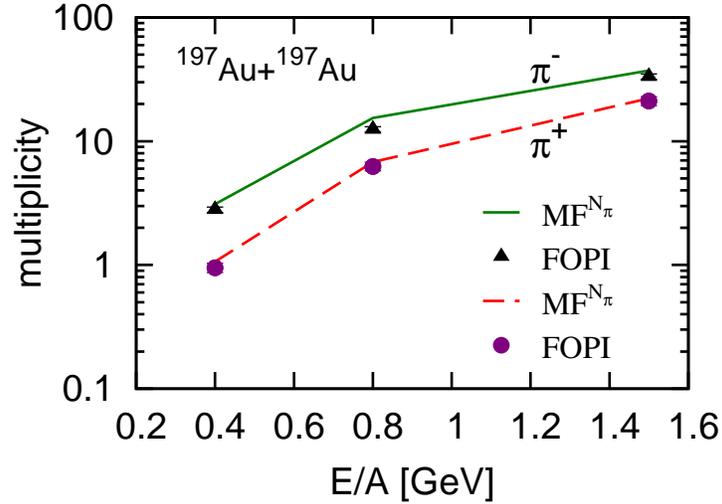}
\caption{(Color online) Pion multiplicity in central Au+Au collisions, as a function of beam energy. Symbols represent data of the FOPI collaboration \cite{FOPI-1}, while lines represent the pBUU calculations with the $N_{\pi}$-adjusted momentum-dependent MF.}
\end{figure}

In testing the characteristics of the $N_{\pi}$-adjusted MF, we examine the momentum dependence of optical potentials in zero-temperature matter. For the optical potential $U^{opt}(p)$, we employ in the relativistic context the following definition:
\begin{equation}
U^{opt}(p)=\epsilon(p)-m-T(p)
\end{equation}
In the equation, $\epsilon(p)$ is the single particle energy corresponding to momentum $p$, and $T(p)$ is the kinetic energy. 

In Fig. 3 we plot the optical potentials for the two parameterizations, as a function of momentum, with different lines representing different indicated densities. The dashed and solid lines represent, respectively, optical potentials from $v_2$-optimized and $N_{\pi}$-adjusted MFs. The momentum dependence in $N_{\pi}$-adjusted MF is indeed softened, consistently with expectation developed on the basis of Fig. 1. 

In \cite{PD-4}, the momentum dependence of the optical potential from (6)-(8) was compared to that found for potentials from microscopic calculations including those relying on the Urbana v14 two-body interaction combined with model VII three-body interaction, i.e. UV14+UVII \cite{UV-AV}, AV14+UVII \cite{UV-AV}, as well as DBHF \cite{DBHF-1,DBHF-2}, BBG \cite{BBG-1,BBG-2} and UV14+TNI \cite{UV-AV}. Regarding those microscopic calculations, the $N_{\pi}$-adjusted MF produces optical potentials closest in form and values to UV14+UVII, with the respective comparison illustrated in Fig. 4. Similarly to \cite{UV-AV}, we compare  the single-particle energy, $\epsilon(p,\rho)-m(\rho)$ between our potential and UV14. The momentum dependence in this representation is implicit. 

\begin{figure}
\centering
\includegraphics[scale=1.5]{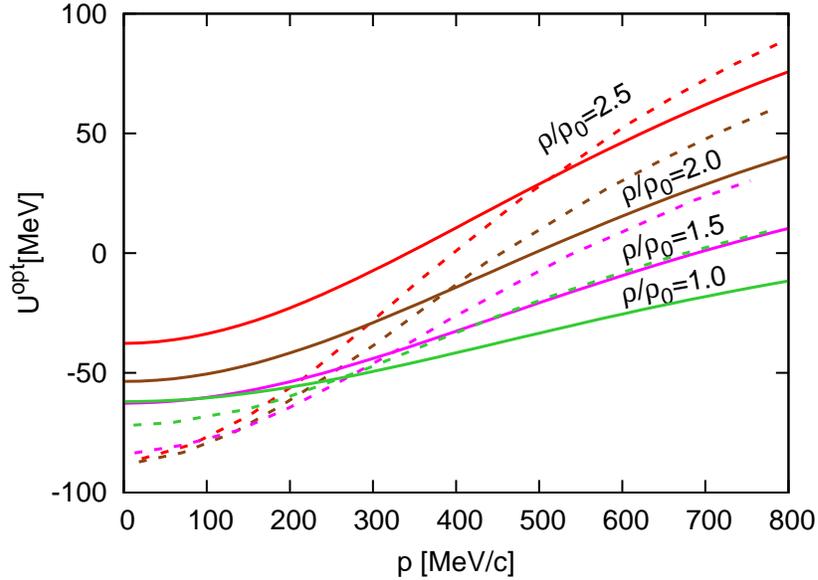}
\caption{(Color online) Optical potential in nuclear matter at different indicated densities, as a function of momentum. Dashed and solid lines represent, respectively, the $v_2$-optimized and $N_{\pi}$-adjusted MFs.}
\end{figure}

\begin{figure}
\centering
\includegraphics[scale=1.5]{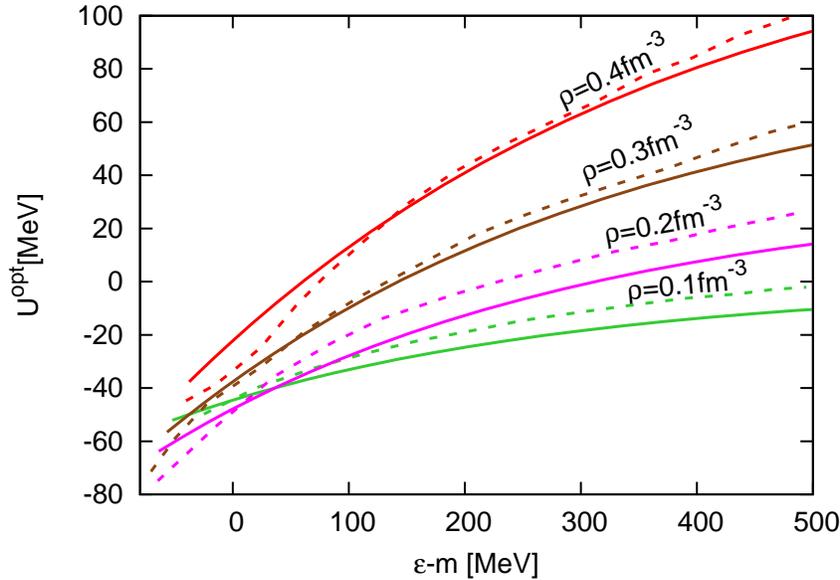}
\caption{(Color online) Optical potential in nuclear matter at different indicated densities, as a function of nucleon energy. Dashed and solid lines represent, respectively, UV14+UVII variational calculations and our $N_{\pi}$-adjusted MF.}
\end{figure}

\subsection{Elliptic Flow}
In the past, anisotropies of collective flow and, in particular, the elliptic flow, were used to test the characteristics of MF momentum-dependence in collisions \cite{PD-4}. The elliptic flow is defined as the second-order Fourier coefficient of azimuthal angle with respect to reaction plane at midrapidity: $v_2=<$cos$(2\phi)$$>$. 

Obviously when more constraints are placed on an MF, such as of the proper description of total pion yields, the description of the measured elliptic flow cannot be generally as good as that achievable without those additional constraints. Figure 5 shows the out-of to in-reaction-plane ratio, $R=\frac{1-v_2}{1+v_2}$, for protons emitted at midrapidity from mid-peripheral Bi+Bi collisions at 400A MeV, as a function of proton transverse momentum. The stronger the elliptic flow, the larger the deviation of R from 1. The filled triangles in Fig. 5 represent the data of the KaoS \cite{KAOS} collaboration, while the dashed and solid lines represent, respectively, the pBUU calculations with $v_2$-optimized and $N_{\pi}$-adjusted MF. The two calculations describe about equally well the KaoS data at intermediate momenta, but the $v_2$-optimized MF is far superior at high momenta.

The difficulty in the simultaneous description of high-momentum $v_2$ and near-threshold pion yields is puzzling and likely points to some limitation in our MF parameterization. One possibility is the lack of anisotropy in the momentum dependence, for anisotropic momentum distributions $f$, when employing (6)-(9). While our implementation (6)-(9) of the MF momentum dependence allows in practice for a higher precision of calculations, than other MF parameterizations \cite{Christian}, it may turn out to be a handicap here. We already undertook steps, cf. the work of Simon and Danielewicz \cite{Christian}, towards implementing anisotropy without compromising calculational precision or speed.
\begin{figure}
\centering
\includegraphics[scale=2.]{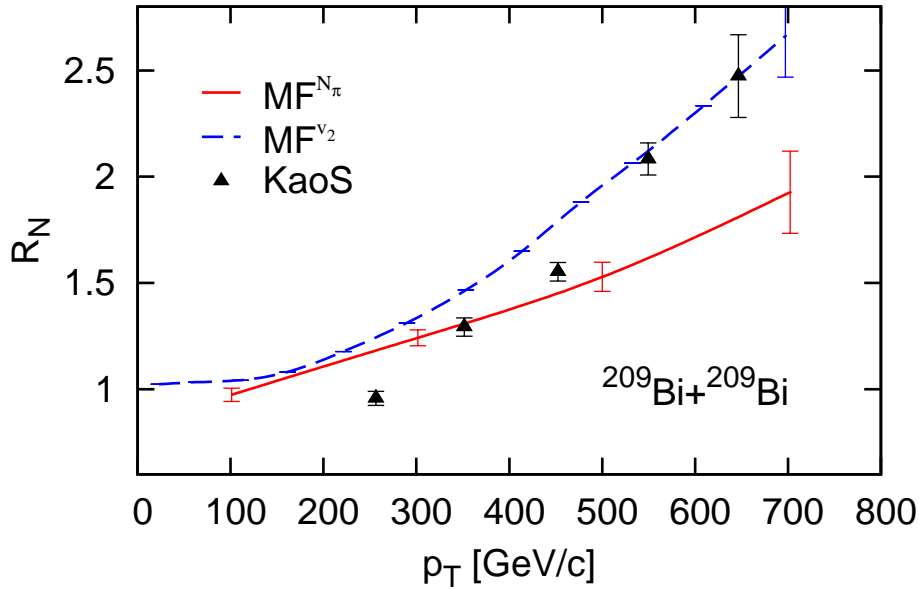}
\caption{(Color online) Ratio of out of reaction plane to in-plane proton yields, as a function of transverse momentum. Symbols represent data from the measurements of the KaoS Collaboration of mid-peripheral Bi+Bi collisions at the beam energy of 400A MeV ($b\simeq8.7fm$) \cite{KAOS}. Solid line represents pBUU calculations with the $N_{\pi}$-adjusted momentum-dependent MF and dashed line represents calculations with $v_2$-optimized momentum-dependent MF. The indicated theoretical errors are statistical, associated with the Monte-Carlo sampling in the transport calculations.} 
\end{figure}

\section{Nuclear Symmetry Energy}
\subsection{Charged Pion Ratios}
Pion observables in HIC are also very important for constraining the stiffness of symmetry energy. Li was first to propose that charged pion ratio is a sensitive observable for high density behavior of symmetry energy \cite{Baoan-ratio}. The link between the pion yield ratio and symmetry energy turned out subsequently to be less straight forward than first proposed \cite{Baoan-ratio}, though, with different transport models contradicting each other as is, in particular, illustrated in Fig. 6.

The isospin contribution to the energy $E_T$ in (5) for pBUU is 
\begin{equation}
E_T=4\int d\vec{r} \ S(\rho) \ \frac{\rho_T^2}{\rho}
\end{equation}
where $\rho_T=\sum\limits_X \rho_X \ t_{3X}$ and $t_{3X}$ is the third component of isospin for species X. The symmetry-energy factor S can be conveniently decomposed as 
\begin{equation}
S(\rho)=S_{kin0}\left(\frac{\rho}{\rho _0}\right)^{\frac{2}{3}}+S_{int}(\rho),
\end{equation} 
where the first r.h.s. term, with $S_{kin0}\simeq12.3$MeV, represents the symmetry energy in absence of interactions, due to Pauli principle, and the second term represents interaction contribution. In \cite{PD-4} and the calculations here so far, the interaction contribution was of the simplest possible linear form
\begin{equation}
S_{int0}(\rho)=S_{int0}\left(\frac{\rho}{\rho _0}\right).
\end{equation}
However, this can be modified to a power parameterization
\begin{equation}
S_{int0}(\rho)=S_{int0}\left(\frac{\rho}{\rho _0}\right)^{\gamma},
\end{equation}
for more generality.
Larger values of $\gamma$ produce symmetry energies rising quickly with density around $\rho_0$. Such symmetry energies are generally termed stiff. Low values of $\gamma$ yield symmetry energies changing slowly around $\rho_0$. These are termed soft. Description of nuclear masses requires $S_{int0} \sim 20$MeV, best accompanied by a positive correlation between $S_{int0}$ and $\gamma$.

\begin{figure}
\centering
\includegraphics[scale=2.]{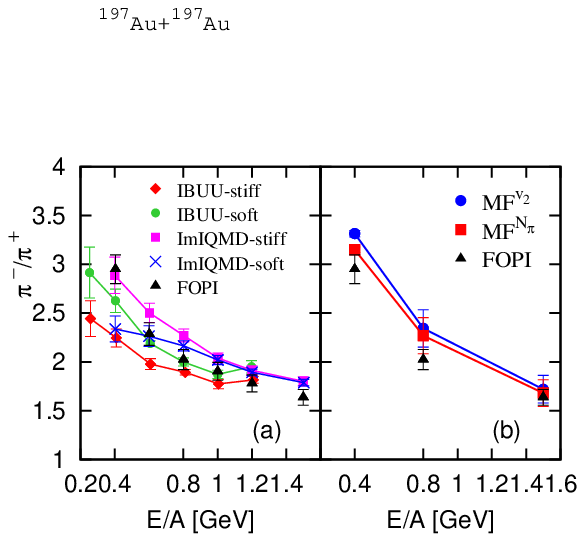}
\caption{(Color online) Pion ratios in central Au+Au collisions, as a function of beam energy. Data of the FOPI Collaboration are represented by filled triangles. The left panel compares predictions from IBUU and ImIQMD models to the data. The IBUU calculations employing stiff and soft symmetry energies are represented there by filled diamonds and filled circles, respectively. The ImIQMD employing stiff and soft symmetry energies are, on the other hand, represented by filled squares and crosses, respectively. The right panel compares predictions from pBUU model to the data. Calculations employing $v_2$-optimized MF and $N_{\pi}$-adjusted MF there are represented there by filled circles and filled squares, respectively. In our calculations here, the potential part of symmetry energy is linear in density.}
\end{figure}

Figure 6 displays ratios of net yields of charged pions stemming from central Au+Au collisions at different beam energies. The filled triangles represent measurements of the FOPI Collaboration \cite{FOPI-2}. Other symbols represent results of different transport calculations. In the panel (a) of Fig. 6, it is seen that, within IBUU calculations \cite{IBUU}, a stiff symmetry energy gives rise to a lower $\pi^-/\pi^+$ ratio than does a soft energy. However, the converse is true for the ImIQMD calculations \cite{ImIQMD},  as seen in the same panel, which is the current contradiction in the literature, mentioned before.

In our own calculations, the $\pi^-/\pi^+$ net yield ratio is practically independent of the details in the momentum dependence of MF, as illustrated in panel (b) of Fig. 6, where we show results utilizing both $v_2$-optimized and $N_{\pi}$-adjusted MF. The results are obtained for Au+Au collisions at $b<2$fm. We use here the linear $S_{int}$, Eq. (12), and either set of results agrees, in practice, with the FOPI measurements. Importantly, we further find that the net charged pion ratio and the agreement with the measurements remain largely independent of the stiffness of symmetry energy. That is illustrated in Fig. 7, where we show pBUU results obtained in calculations of central Au+Au collisions at 200 and 400A MeV, when changing $\gamma$ in the symmetry energy (13).

One detail in pBUU that may give rise to different sensitivity to the symmetry energy for net pion yields, than in other transport calculations, is the presence of a strong interaction potential acting on pions and driven by isospin imbalance. From (10), that potential is 
 \begin{equation}
 U_{\pi^{\pm}}=\mp 8 \: S_{int0} \: \rho_{T} \: \frac{\rho^{\gamma-1}}{\rho_0^{\gamma}}
 \end{equation}
In IBUU and ImIQMD, strong-interaction potentials acting on pions are lacking.

\begin{figure} 
\centering
\includegraphics[scale=1.5]{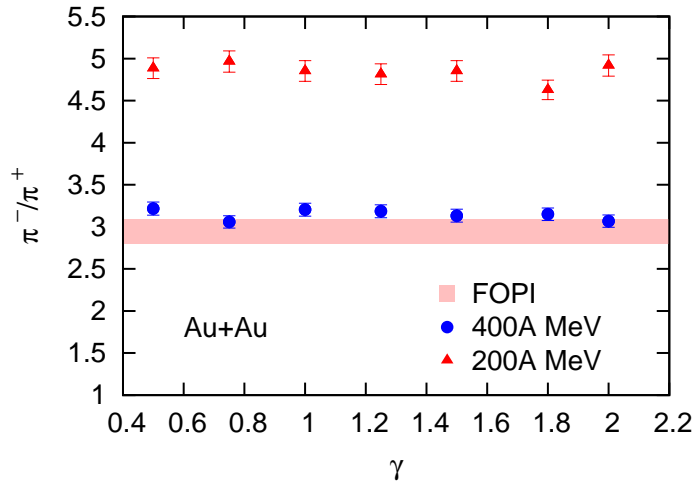}
\caption{(Color online) Ratio of net charged pion yields in central Au+Au collisions at 400A MeV and 200A MeV, as a function of the stiffness of symmetry energy $\gamma$, from pBUU calculations using $N_{\pi}$-adjusted MF. The dashed region represents the 400A MeV FOPI measurement. The theoretical errors are due to statistical sampling in the pBUU calculations.}
\end{figure}

Pion-nucleus optical potential has been used to explain the existence of pionic atoms. Toki et al., in particular, constructed a pion potential that successfully described the deeply bound states of pionic atoms \cite{Toki}. In Fig. 8, the potential in pBUU, for three values of $\gamma$, is compared to that of Toki, for $^{197}$Au. Given that our potential in the form (14) can only represent the so called s-wave contribution to the $\pi$-nucleus potential, we drop, in the comparison, the small p-wave contribution to the potential of \cite{Toki}. The tails are different in our potentials compared to Toki, due to excessively abrupt changes of density in the semiclassical Thomas-Fermi model (the T=0 limit of our transport model) in the surface region. For pions moving across a HIC zone, however, the most important is the magnitude of the potential over regions where density changes slowly, including nuclear interior in the ground state. In the interior, our potentials for $\gamma$ from 1 to 2 are within $30\%$ away from the Toki's potential.

The potentials of different sign for $\pi^+$ and $\pi^-$, each equal in magnitude to the difference between neutron and proton mean fields, and also a difference in the potentials for $\Delta$, may produce enough difference in the propagation of charged pions in the pBUU and other models to affect predictions.

\begin{figure}
\centering
\includegraphics[scale=1.5]{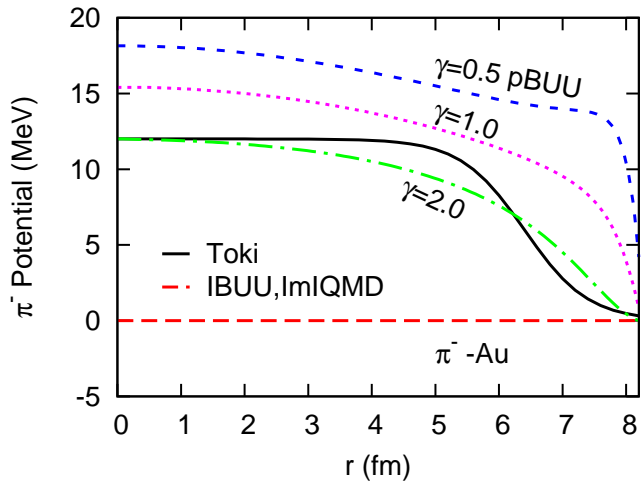}
\caption{(Color online) S-wave contribution to $\pi-^{197}Au$ optical potential. Solid line represents the work of Toki \cite{Toki}. Short-dash, dotted and dash-dotted lines represent pion potentials from pBUU parameterization for $\gamma=0.5, 1.0, 2.0$, respectively, in the interaction part of the symmetry energy. Long dashed line represents the lack of corresponding potentials in the IBUU and ImIQMD models. }
\end{figure}

\subsection{Differential Pion Ratios}
While we found no sensitivity of net charged pion yield ratios in pBUU, around threshold, to $S(\rho)$, still the general idea \cite{Baoan-ratio} contains convincing elements. Potentially, more differential ratios of charged pion yields could provide access to $S(\rho)$ at supranormal densities.

In Fig. 9-11, we explore the sensitivity of charged-pion spectra to the stiffness of symmetry energy. The first two figures illustrate the $\pi^-/\pi^+$ ratio as a function of pion c.m. energy and the third illustrates the average c.m. energies for the charged pions. Difference in the average c.m. energies, between $\pi^+$ and $\pi^-$, is additionally plotted in Fig. 12, as a function of the stiffness $\gamma$ of the symmetry energy, for Au+Au at 200 MeV/nucleon.

\begin{figure}
\centering
\includegraphics[scale=1.5]{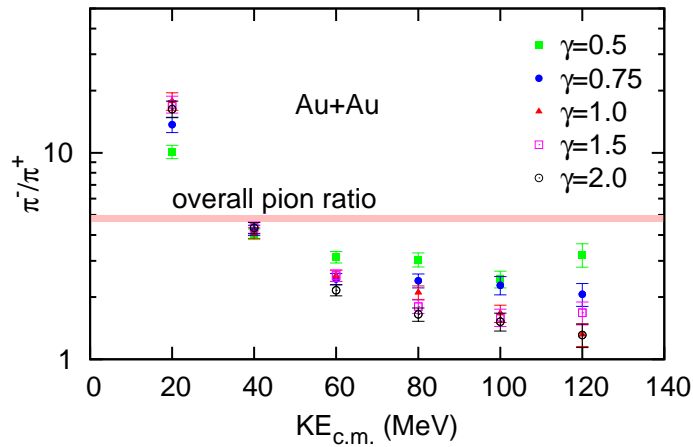}
\caption{(Color online) Charged pion ratio in central Au+Au collisions at 200A MeV, as a function of kinetic energy in the center of mass frame, for different values of the stiffness $\gamma$ of the symmetry energy, from 0.5 to 2.0. The horizontal line represents the ratio of net charged pion yields.}
\end{figure}

\begin{figure}
\centering
\includegraphics[scale=1.5]{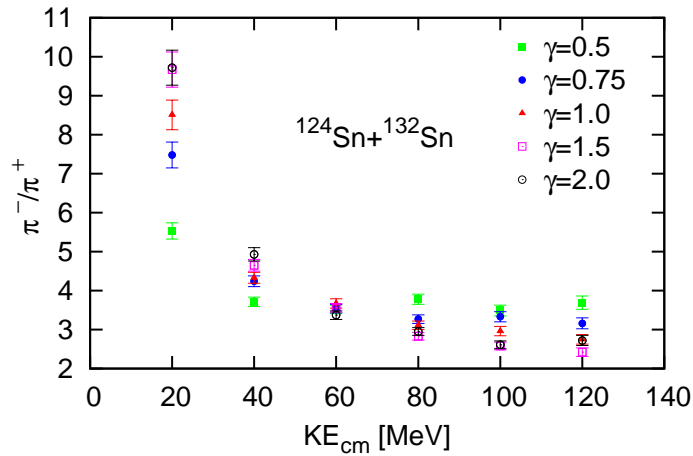}
\caption{(Color online) Charged pion ratio in central $^{124}$Sn+$^{132}$Sn collisions at 300A MeV, as a function of kinetic energy in the center of mass frame, for different values of the stiffness $\gamma$ of the symmetry energy, from 0.5 to 2.0.}
\end{figure}

\begin{figure}
\centering
\includegraphics[scale=1.5]{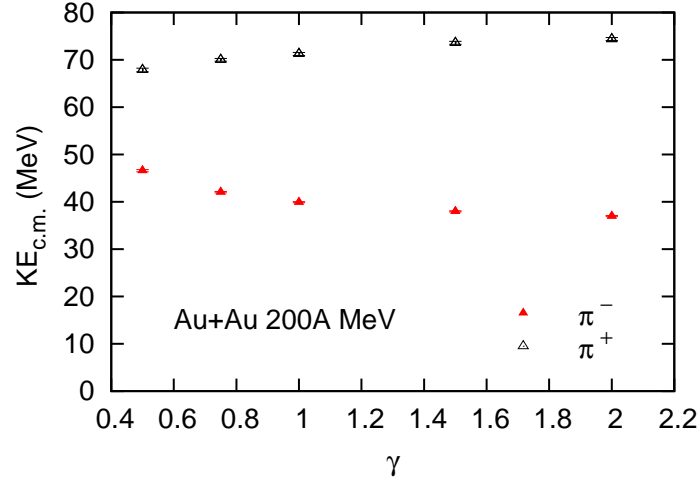}
\caption{(Color online) Average center-of-mass kinetic energy of $\pi^+$ and $\pi^-$ in central Au+Au collisions at 200A MeV, plotted against stiffness $\gamma$ of the symmetry energy.}
\end{figure}

\begin{figure}
\centering
\includegraphics[scale=1.5]{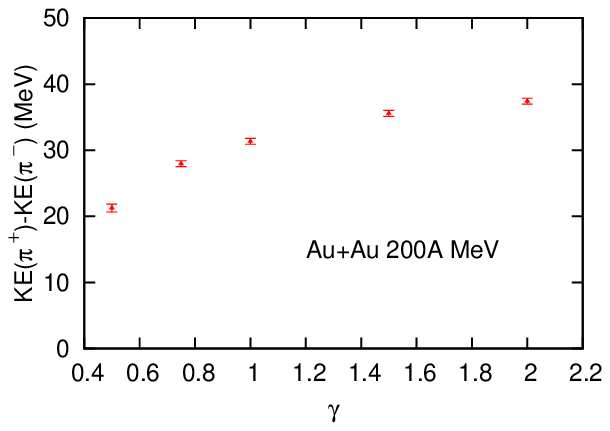}
\caption{(Color online) Difference between average c.m. kinetic energy of $\pi^+$ and $\pi^-$ in central Au+Au collisions at 200A MeV, plotted against stiffness $\gamma$ of the symmetry energy.}
\end{figure}

The figures display competing effects of the isospin content of the system, of Coulomb interactions and of symmetry energy. Obviously, the neutron excess generally makes negative pions more abundant than positive, with the effect amplified by larger isospin for the pions than nucleons. The long-range Coulomb interactions play the primary role in making the $\pi^-/\pi^+$ ratio dependent on the energy of the emitted pions. Thus, after the pions cease to interact strongly and move out from the reaction region, described then by primordial spectra sharing to a degree characteristics between $\pi^+$ and $\pi^-$ (and $\pi^{\circ}$), the Coulomb interactions accelerate $\pi^+$ and decelerate $\pi^-$. The relative Coulomb push boosts the $\pi^-/\pi^+$ ratios at low c.m. energies, above the overall ratio for the reactions, and lowers the ratios at high c.m. energies, see Figs. 9 and 10. The push also gives rise to substantially higher average c.m. energies for $\pi^+$ than $\pi^-$, see Figs. 11 and 12.

Contributions to mean-field potentials associated with the symmetry energy principally act opposite to Coulomb interactions, but they act there where pions continue to rescatter, in fact with large cross-sections due to the formation of $\Delta$-resonance, down to low densities. The scattering tends to erase the impact of different accelerations for $\pi^+$ and $\pi^-$ (and for nucleons and $\Delta$'s with different isospin as well) due to the isospin-dependence of mean fields. With the scattering rates being linear in density, the mean fields can win over the rescattering, in the low density region, if their dependence on density is slower than linear. The low-energy part of the spectrum is generally dominated by particles emitted from lower density regions, late in the history of the reactions. In Figs. 9 and 10, we can see that the symmetry energy is indeed effective in countering the effects of Coulomb enhancement of the low-energy $\pi^-/\pi+$ ratio when $\gamma \lesssim 1$ and the interaction symmetry energy is large at low densities. At $\gamma > 1$, the effect fizzles out. Notably excitation of the medium suppresses the role of Pauli principle and of the associated kinetic contribution to the symmetry energy. In Fig. 11, we can see that the impact of the stiffness of symmetry energy, on $\pi^+ - \pi^-$ average-energy difference, weakens past $\gamma \approx$ 1.

With regard to the particles emitted at higher c.m. energies, that tend to stem from early stages of the reaction and higher densities, another high-density effect of the symmetry energy comes into play. Namely, a stiff symmetry energy pushes away the neutron-proton asymmetry from the high-density region [25], see Fig. 13. With the reduction in the high-density asymmetry, the $\pi^-/\pi^+$ ratio gets reduced at high c.m. energies. Thus, qualitatively a stiff symmetry energy acts in this energy region as the relative Coulomb boost, cf. Figs. 9 and 10. With this, it becomes possible to access the stiffness of high-density symmetry-energy through the high-energy $\pi^-/\pi^+$ yield ratio.

\begin{figure}
\centering
\includegraphics[scale=1.5]{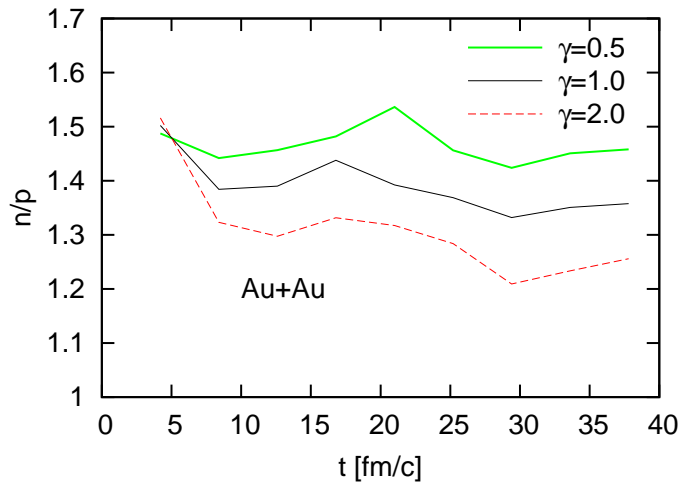}
\caption{(Color online) Ratio of neutron-to-proton numbers at net supranormal densities, $\rho > \rho_0$, in central Au+Au collisions at 200A MeV, as a function of times. At early times, the numbers in the ratio are marginal, and the ratio, thus, not very meaningful.}
\end{figure}

In the earlier version of this work [28], we also explored the $\pi^-/\pi^+$ yield ratio in the direction out of the reaction plane as a probe of the symmetry energy at supranormal densities. In that direction the high-density matter is directly exposed to the vacuum. However, with a higher statistics in the calculations, our claimed directional signal for the symmetry energy has weakened.

\section{Conclusion}
With a new parameterization for momentum-dependent MF, pBUU gives a reasonable description of pion multiplicities in moderate-energy central HIC. The puzzling finding is that the same parameterization of the MF momentum-dependence cannot be simultaneously used for describing the net pion yields around threshold and the high-momentum elliptic flow of protons. One potential avenue for resolving this puzzle is in the need for an anisotropy in the MF momentum dependence, when the underlying momentum distribution is anisotropic. We compared our new momentum dependence of nucleonic optical potential with several microscopic calculations. The modified potential is within the realm of uncertainties for microscopic predictions, just like the previous potential.

Next, we used pion ratio observables to study the symmetry energy behavior at higher density than normal. While IBUU and ImIQMD yield opposing sensitivities to the density dependence of symmetry energy, for $\pi^-/\pi^+$ net yield ratios, we find no significant sensitivity for that ratio to $S(\rho)$ in pBUU. One factor affecting that sensitivity may be the pion optical potential in pBUU, driven by isospin asymmetry. We examined the dependence of charged pion ratio on pion c.m. energy. To isolate the effect of symmetry energy at supranormal densities, we looked at the high energy tail of the spectra---there a clear sensitivity of pion ratio to different forms of supranormal symmetry energy is seen. Additionally, the difference of average c.m. kinetic energy of emitted $\pi^+$ and $\pi^-$ also shows a distinguishing power for different symmetry energies. In the first version of this paper, we applied combined energetic and angular cuts to the pion ratios and proposed it as a new differential observable for future experiments, Ref. \cite{Jun}.

\section*{Acknowledgement}
The authors would like to thank Betty Tsang and Bill Lynch for their insightful suggestions, and also thank Brent Barker for helpful discussions. Some early calculations of the pion production around threshold we carried out with pBUU by Michelle Mosby. This work was supported by the U.S. National Science Foundation under Grant PHY-1068571.

\end{document}